\ifx\pdfoutput\undefined        
  \documentclass[twocolumn]{article}
  \usepackage{url}
\else                           
  \documentclass[pdflatex,twocolumn]{article}
  \usepackage[bookmarks=false]{hyperref}
\fi

\usepackage{graphicx}
\usepackage{times}

\newcommand{\ysci}[5]{ (#1) #5, {\em Science }{\bf #2}, #3--#4.}
\newcommand{\ybook}[3]{ (#1) {\em #2}, #3.}

\begin{document}

\title{Exploring the Galaxy using space probes}
\author{R.\ Bj\o{}rk$^1$}
\date{\small
$^1$Niels Bohr Institute, Juliane Maries vej 32, DK-2100 Copenhagen
\O, Denmark, e-mail: rbjk@astro.ku.dk}

\maketitle

\abstract{This paper investigates the possible use of space probes
to explore the Milky Way, as a means both of finding life elsewhere
in the Galaxy and as finding an answer to the Fermi paradox. I
simulate exploration of the Galaxy by first examining how long time
it takes a given number of space probes to explore 40,000 stars in a
box from -300 to 300 pc above the Galactic thin disk, as a function
of Galactic radius. I then model the Galaxy to consist of
$\sim{}260,000$ of these 40,000 stellar systems all located in a
defined Galactic Habitable Zone and show how long time it takes to
explore this zone. The result is that with 8 probes, each with 8
subprobes $\sim{}4\%$ of the Galaxy can be explored in
$2.92\cdot{}10^{8}$ years. Increasing the number of probes to 200,
still with 8 subprobes each, reduces the exploration time to
$1.52\cdot{}10^{7}$ years.}

\section*{Introduction}
Exploring the Galaxy by the use of interstellar space probes is the
only way to perform detailed investigations of extrasolar planets
believed to harbor life.

Also the Fermi paradox, \begin{quote} \emph{If there are
extraterrestrial civilizations out there then where are they? Why
haven't we seen any traces of intelligent extraterrestrial life,
such as probes, spacecraft or transmissions?}
\end{quote} can be investigated if we know how long time it
takes to explore the Galaxy with space probes.

Here I simulate exploration of the Galaxy with space probes using a
numerical model of the Galaxy. But before going into detail with the
model, I describe the initial parameters needed to built a model of
the Milky Way.

In Section \ref{Initialparameters} the initial parameters for the
simulation is derived. Section \ref{SectionExploring40000stars}
presents a model for exploring 40,000 stars in a box covering a
distance of 300 pc above and below the Galactic plane as a function
of Galactic radius. Section \ref{SectionModelingthewholeGalaxy}
models the whole Galaxy based on these results and presents the
final results and discussion.

\section{Initial parameters}\label{Initialparameters}
There are roughly 100 billion stars in the Milky Way (Allan 1973).
The fraction of these stars that have planets suitable for harboring
life is estimated to be $\simeq{} 10 \%$, all located in the so
called Galactic Habitable Zone (GHZ), which spans the area $3-11$
kpc from the Galactic Center (Lineweaver \emph{et al.} 2004). This
GHZ is comparable with the one found by Pe\~na-Cabrera et al.
(2004), which extents from $4-17.5$ kpc. In the following
calculations, the Lineweaver GHZ is used.

There are roughly 820 F, G, K and M stars pr $10^{4}$ pc$^{3}$ in
the solar neighborhood, equal to a square box of $21.54^3$ pc$^{3}$
(Allan 1973). This corresponds to 0.0615 stars per pc$^{3}$ if it is
assumed that $\sim{}50\%$ of the stars are in binaries and a binary
is counted as one star. O and B stars are disregarded as these have
to short a lifetime too harbor planets where life has time to
evolve. Spectral class A stars are included as it is argued by
Pe\~na-Cabrera et al. (2004) that these have a long enough lifetime
to allow at least simple life to evolve. Binary systems consisting
of two low mass stars are not disregarded as simulations have shown
that these can harbor terrestrial-like planets (Lissauer \emph{et.
al.} 2004)

The stellar number density of the Milky Way disk is assumed to
decline exponentially with distance from the Galactic Center (GC).
Measurements of the scale length gives results in the range 2.5-3.5
kpc (see Sackett 1997 for a review). Star counts from the Two-Micron
All-Sky Survey (2MASS) yield 3.3 kpc (López-Corredoira et al. 2002)
whereas direct measurements of M stars with the Hubble Space
Telescope by Zheng et al. (2001) results in 2.75 kpc. In the
following a scale length of 3 kpc is used.

The area to be explored is defined as the area between 3 kpc and 11
kpc from the GC and 300 pc above and below the Milky Way's thin
disk, as this last figure is the scale-height of the Galactic thin
disk (Gilmore \& Reid 1983).

To find the number of stars in this area the stellar number density
profile is simply integrated over this interval, with binaries
counted as one star. With $z$ denoting the height respectively above
and below the Galactic Plane, $r$ denoting the radial distance from
the GC and $\sigma{}_{GC}$ denoting the stellar number density at 3
kpc radial distance from the GC, the number of habitable stars,
$N_{hab}$, is
\begin{eqnarray}
N_{hab}=\sigma{}_{3\textrm{\tiny{kpc}}}\int_{3\;\textrm{\tiny{kpc}}}^{11\;\textrm{\tiny{kpc}}}\hspace{-.5cm}2\pi{}r\cdot{}e^{\frac{-r}{3\;\textrm{\tiny{kpc}}}}\;dr\cdot{}2\cdot{}\int_{0\;\textrm{\tiny{kpc}}}^{0.3\;\textrm{\tiny{kpc}}}\hspace{-.5cm}e^{\frac{-z}{0.3\;\textrm{\tiny{kpc}}}}\;dz~,
\end{eqnarray}
equal to $1.17\cdot{}10^{10}$. The factor of two is caused by the
equality between $-z$ and $z$.

These $1.17\cdot{}10^{10}$ habitable stars are an overwhelmingly
large amount, and too many to include in a simulation. My approach
will therefore be to examine how long time it takes a number of
probes to explore a small number of stars, and see how this
exploration-time scales as a function of the stellar number density,
i.e. the Galactic radius. I also investigate how the exploration
time scales as a function of the number of probes used. These
results are then later used when simulating exploration of the
entire Galaxy.

\section{Exploring 40,000 stars} \label{SectionExploring40000stars}
In this simulation, I investigate how long it will take to explore a
volume containing 40,000 stars. In the following ``stars'' designate
stars that can possibly sustain life. Thus, as already mentioned, O
and B stars are excluded in the models.

\subsubsection{The model}
In the first part of my simulation 3D box-systems are modeled, each
containing 40,000 stars, but located at different Galactic radii.
The already mentioned exponential decline of the stellar density
with Galactic radius means that there will be fewer stars per cubic
pc further away from the GC, and each box will thus become bigger in
order to contain the 40,000 stars. All the boxes go from -300 to 300
pc in the $z$ direction. In regards to the exponential decline of
the stellar density any spiral structure that the Galaxy might have
is disregarded, as I assume that the density of low-mass stars does
not increase strongly in the spiral arms.

\subsubsection{Exploration algorithm}
The 40,000 stars are explored by sending out one host probe which
travels to some faraway star referred to as the ``destination
star''. Once the probe arrives, it dispatches a number (4 or 8)
smaller probes, that in total investigate the 40,000 nearest stars.
They do this by always moving to the star nearest to their current
location that have not been explored already. The distance and
position of this star can easily be determined from its parallax.
After all the 40,000 stars have been explored the probes return to
the destination star, where they dock with the host probe for
maintenance and prepare to travel to a new destination star.

The time it takes to travel back to the host probe from the last
star explored, once all 40,000 stars are covered, is a tiny fraction
of the time it takes to explore the 40,000 stars, and thus this
return time has no impact on the total exploration time.

Once all the 40,000 stars have been explored and the probes have
returned to the host probe, this moves to a new destination star
some distance away where the smaller probes are released again to
explore the 40,000 nearest stars and so on. In this way, the whole
galaxy can be covered in a finite amount of time.

For all probes a speed of $0.1c$ is assumed. This velocity is low
enough that effects due to general relativity can be ignored, yet
high enough that the travel time between stars are on the order of
years. The smaller probes only do flyby investigations of the stars
since the time and energy required to first brake and then
reaccelerate the probe would slow the exploration time considerably.
The time required to brake the spacecraft, examine all the planets
in the given system in detail, and reaccelerate the spacecraft will
be at least of the same order of magnitude as the travel time
between two stars, thus reducing the number of stars visited in a
given amount of time by at least a factor of 2. Therefore flyby
investigation is assumed.

The purpose of the flyby spaceprobes is to detect whether there are
any possible habited planets in the system. If this is the case then
a new probe can be launched from the mother planet to do detailed
investigation. It should be relatively easy to detect life on the
surface of planets even from a flyby spaceprobe. To use our own
civilization as an example, the light-pollution and heavy
radio-broadcasted communication would reveal our presence on the
Earth easily to any flyby probe entering the Solar System, even if
it were only equipped with technological equipment available today,
such as a telescope equal to the Hubble Space Telescope's and a
radio receiver.

\begin{figure}[!ht]
   \centering
   \includegraphics[width=\columnwidth]{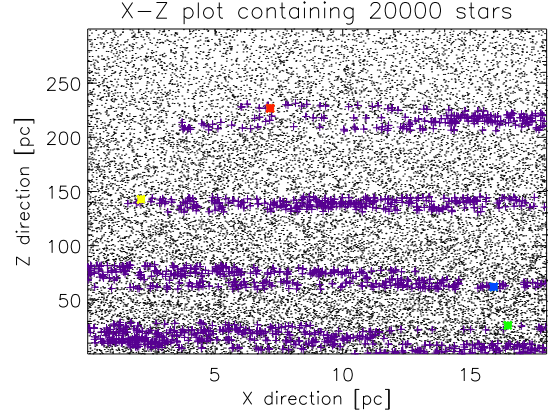}
      \caption{\em This figure shows the $x$-$z$ plane of a box containing 20,000 stars
      from 0 to 300 pc in the $z$-direction and at 3,000 pc from the GC.
      4 probes, each marked by their own color at their present location, are used to explore
      the stars. Unexplored stars are marked with a black dot and stars that have been visited are marked by a purple $+$.
      At the time of this snapshot 1300 stars  have
      been explored. Notice how the probes are confined to certain z-intervals.}
      \label{galaxy.z.8000.800}
\end{figure}

To prevent the small probes from visiting the same stars twice, each
probe is limited to only investigate stars at a given height from
the disk-plane. Thus the first probe will e.g. only explore stars in
the range $z = 0-50$ pc and so on.  In reality this could be done by
making the probes calculate their distance from the disk plane based
on parallax to a number of known disk plane stars.
This approach effectively prevents probes from exploring the same stars. \\
The exploration pattern of such a setup can be seen in Fig.
\ref{galaxy.z.8000.800}.

\begin{figure}[!ht]
  \centering
   \includegraphics[width=\columnwidth]{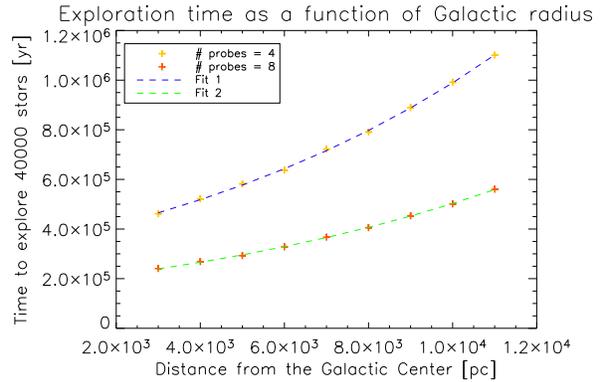}
      \caption{\em Exploration time of 40,000 stars from -300 to 300 pc in the z-direction as a
      function of Galactic radius. The fitted functions are both
of the form $a\cdot{}e^{b\cdot{}r}$.} \label{plot-radius-time}
\end{figure}

The time it takes a given number of probes to fully explore a 3D box
containing 40,000 stars as a function of Galactic radius is given in
Fig. \ref{plot-radius-time}. The two different colored symbols are
if either 4 or 8 small probes are used. The fitted functions are
both exponential functions of the form $a\cdot{}e^{b\cdot{}r}$. For
4 probes $a=334,636\; \textrm{[yr]}$ and $b=1.07732\cdot{}10^{-4}\;
\textrm{[pc$^{-1}$]}$ and for 8 probes $a=172,734\; \textrm{[yr]}$
and $b=1.05971\cdot{}10^{-4}\; \textrm{[pc$^{-1}$]}$. It thus takes
of the order of $10^5$ years to explore 40,000 stars, covering $z$
from -300 to 300 pc.

Having found out how long time it takes to explore 40,000 stars, as
a function of Galactic radius, this result is used to simulate
exploration of the entire Galaxy.

\section{Modeling the whole
Galaxy}\label{SectionModelingthewholeGalaxy} The problem with
modeling the whole Galaxy is that even when this is modeled
consisting of systems each with 40,000 stars, there are still
$\sim{}260,000$ such system in the Galaxy. This is a bit too many to
be handled in a reasonable amount of time, so I focus only on the
time needed to explore our quadrant of the Galaxy. Thus, only an
area one fourth the size of the total circumference of the Galaxy,
centered on the Solar System is explored. This reduces the number of
systems needed to be explored by a factor of 4. In this quadrant of
the Milky Way there will thus be $\sim{}65,000$ systems consisting
of 40,000 stars each, clearly illustrating just how enormous the
Galaxy is. In the following ``stars'' mean systems of 40,000 stars.

The model of the Galaxy is build so that the stellar density
declines exponentially with Galactic radius as already explained in
Sec. \ref{Initialparameters}.

Each host probe always moves to the star located nearest to its
present location that has not been explored yet. This means that not
all stars in the vicinity of the home planet will be covered, but it
maximizes the number of stars explored as a function of time.

To prevent the host probes from exploring the same stars each host
probe is assigned an angle in the disk plane as seen from the mother
planet if looking towards the GC, which it has to stay inside. For
example, if there are 4 probes, then the first could be assigned an
angle from 0 to $\pi{}/2$ and so on. This can easily be accomplished
in reality by telling the probe to fix its orientation by means of
distant galaxies. In my simulations the angle is chosen such that
the number of stars are the same in each angle. This gives a larger
angle to the probes exploring the outer regions of the Galaxy
compared to the probes in the inner part of the Galaxy.

\subsection{Results}
I now simulate the entire Galaxy from 3 kpc to 11 kpc and explore
one quarter of it. 4 or 8 host probes is send out. Each star in the
Galaxy now represents 40,000 stars and each time a probe has
explored one star here, the results from Fig. \ref{plot-radius-time}
is used to add the time needed to explore the 40,000 stars at this
point.

 \begin{figure}[!ht]
  \centering
   \includegraphics[width=\columnwidth]{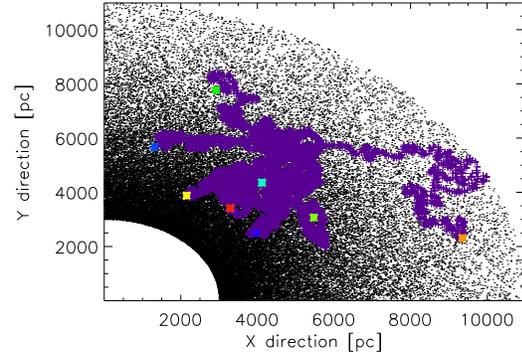}
      \caption{\em A quadrant of the Galaxy, being explored by 8 probes (each with 8
      subprobes). The
      probes, each marked by their own color, start from a distance of 8,000 pc from the GC, and at an angle of
      45$^{\circ{}}$ from the x-axis. The units on both axes are in
      parsec. Explored stars are marked by a purple +.
      Note that more probes are dispatched in the direction of the GC, as the
      number of stars are greater here, and the angle in which the individual probes
      must stay inside thus smaller. Stars inside 3 kpc and outside 11 kpc are not shown, as these are not explored.} \label{galaxy.real.n8.10000}
\end{figure}

\begin{figure}[!ht]
   \centering
   \includegraphics [width=\columnwidth]{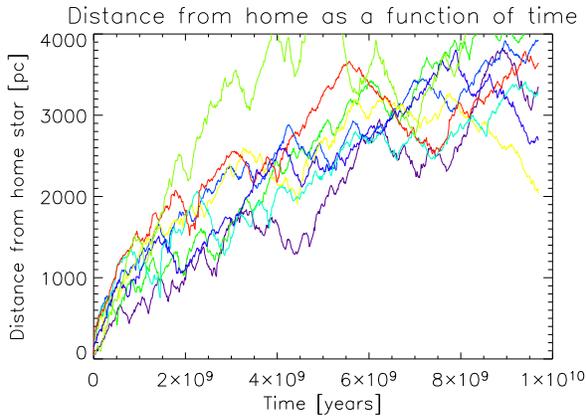}
      \caption{\em This figure shows the distance from the home star, as a function of time, for
      the individual host probes. Each probe is marked by it's own color.
      The probes each manage to a travel distance of almost 4 kpc.}
      \label{plot-time-distfromhome}
\end{figure}

For 4 probes, each with 4 subprobes, the time required to explore
10,000 systems, each containing 40,000 stars, is $1.16\cdot{}10^{9}$
years which is around 10 percent of the age of the Universe
according to observational data of the microwave background
radiation (Spergel et. al. 2006). As this is a completely
unrealistic time for any civilization to wait, the simulation was
also run for 8 probes, each with 8 subprobes. The Galaxy from this
simulation is shown in Fig. \ref{galaxy.real.n8.10000}. Here the
time is $2.92\cdot{}10^{8}$ years, again an immense time. However,
as seen in Fig. \ref{plot-time-distfromhome} the probes manage to
traverse distances of almost 4 kpc during that time, bringing them
to the inner and outer edges of the Galaxy in the quadrant where
they started.

\subsection{Discussion}
The scope of the simulation becomes clear when one realizes that the
above mentioned times are only for exploring 10,000 of the 260,000
systems of 40,000 stars each. Thus it takes 2\% of the age of the
Universe to explore 1/26 $\sim{}$3.85\% of the Galaxy, using 8
probes with 8 subprobes. To explore the entire Galaxy with probes
like this is virtually impossible.

One could argue that using only 8 probes with 8 subprobes each is
totally inadequate for exploring a whole galaxy. This is a good
argument, but it must be remembered that these probes are by no
means similar to the small interplanetary probes that are known
today. They will be much larger and much more expensive to produce
then any space probe previously built since they have to last for
$\sim{}10^{9}$ years and be able to repair themselves. The economic
feasibility of building such probes will not be discussed here.

However, in order to accommodate this criticism, a simulation of the
exploration of the Galaxy, using 200 host probes, each with 8
subprobes was ran. The result is that in this case, it took
$1.52\cdot{}10^{7}$ years to explore the 10,000 nearest systems of
40,000 stars each.

I realize that there are other problems with a simulation of the
above type. First of all, it is assumed that the probes never fail
or are destroyed. It seems very unlikely that a probe would even
function the $\sim{}10^5$ years it takes to explore the 40,000 stars
from -300 to 300 pc in the $z$-direction. All though I have used a
relatively high probe velocity of 0.1$c$, it is a possibility that
even faster probes could be invented, and thus decreasing the
exploration time significantly.

One could also contemplate the idea of launching self-replicating
probes i.e. probes that are able to build copies of themselves by
harvesting materials from each stellar system they pass.

The construction of such probes are technologically as difficult as
producing the conventional probes proposed to be used to explore the
Galaxy, as these conventional probes must operate for millions, if
not billions, of years. Therefore one can argue that
self-replicating probes should instead be used to explore the
Galaxy, as using such probes will lead to much faster exploration
times, as the number of probes increase as time goes by.

In fact if self-replicating probes, or von Neumann probes as they
are also termed, were used to explore the Galaxy it has been shown
that a search of the entire Galaxy will take
$4\cdot{}10^{6}-3\cdot{}10^{8}$ years dependent on the speed of the
probes (Tipler 1980). This is much faster than using the
non-replicative probes proposed in this paper.

However, one should note that there could be complications with
using self-replicating probes. Tipler (1980) himself points out that
the program controlling the self-replicating probes would have to
have so high an intelligence that it might ``go into business for
itself'' and become out of control of the humans who designed it,
resulting in unforeseeable consequences. Since the machines uses the
same resources as humans, a self-replicating machine might regards
humans as competitors and try to exterminate them. Chyba (2005) also
points out that self-replicating probes might evolve to prey on each
other, creating a sort of machine food-chain. This would of cause
drastically reduce their exploration rate.

Therefore the conclusion is that if perfect self-replicating probes
could be built, these could explore the Galaxy much faster than the
probes suggested here. However, building less-then-perfect
self-replicating probes could, in the worst case scenario, have
fatal consequences for the human race.

There is also the problem that on the long time scales used for
modeling the whole Galaxy, the differential rotation of the Galaxy
will have an influence on the exploration time as well as the probes
distance from the home planet and this effect has been neglected.

And finally there is the important consideration that the search
techniques proposed in this paper is basically a blind search. Each
stellar system is searched individually, even though it may not host
any planets at all. Of cause it would be much faster only to search
the stars around which planets were known to exist. If the
likelihood of planets existing around a star as a function of
stellar type were known, one could implement this in a simulation to
find the reduction in time by searching only stars with planets.
Unfortunately we do not know this function and can thus only guess
at the reduction in time. A reasonable assumption would that if only
half the stars have planets, the search time will be reduced by
roughly a factor of two. Note that the search should probably not be
limited to systems with Earth-like planets. We do not know what form
an extraterrestrial lifeform might assume, and thus we cannot
exclude any planetary systems from our search.

In this regard it is worth mentioning the DARWIN space telescope,
currently in development by the European Space Agency. DARWIN, which
is scheduled for launch around 2015, will be able to obtain spectra
of planetary atmospheres of any terrestrial-like planets orbiting
one of the nearest 1000 stars. This mission, together with other
missions such as ``COROT'' or the ``Terrestrial Planet Finder'',
both of which will be able to detect terrestrial-like planets around
the closest stars, will in the near future provide us with candidate
planets to be searched for life. Thus there is a reasonable
possibility that in the nearby future the first human-made
interstellar probes will begin to search the Galaxy for life. Thus
it is important to develop search strategies that will maximize the
exploration potential.

Returning now to the results obtained in this paper, I seem to be
able to conclude based on the results from my simulations that
exploring the Galaxy by sending out probes to visit the other stars
is horribly slow. However, unless travel methods are invented which
gives access to faster-than-light-travel, there seems to be no
alternative way to proceed than this proposed process. This could
offer a possible explanation to the Fermi paradox. We have not yet
been contacted by any extraterrestrial civilizations simple because
they have not yet had the time to find us. Searching the Galaxy for
life is a painstakingly slow process.

\section*{References}

\begin{list}{}{\leftmargin 3em \itemindent -3em\listparindent \itemindent
\itemsep 0pt \parsep 1pt}\item[]

Allan, C.W. \ybook{1973}{Astrophysical Quantities}{London: Athlone
Press, 3.rd ed.}

Chyba, C. F. \& Hand, K. P. (2005), ASTROBIOLOGY: The Study of the
Living Universe, \emph{Annual Review of Astronomy \& Astrophysics},
{\bf{}43}, 31-74

Gilmore, G. \& Reid, I.N. (1983), The vertical structure of the
Milky Way's stellar disk, \emph{MNRAS}, {\bf{}202}, 1025

Lineweaver, C.H.; Fenner, Y.; Gibson, B.K.
\ysci{2004}{303}{59}{62}{, The Galactical Habitable Zone and the Age
Distribution of Complex Life in the Milky Way}

Lissauer J.J., Quintana E.V., Chambers J.E., Duncan M.J., Adams F.C.
(2004), Terrestrial Planet Formation in Binary Star Systems,
\emph{RevMexAA}, {\bf{}22}, 99-103

López-Corredoira M.; Cabrera-Lavers, A.; Garzón, F.; Hammersley, P.
L. (2002), Old stellar Galactic disc in near-plane regions according
to 2MASS: Scales, cut-off, flare and warp, \emph{A\&A}, {\bf{}394},
883-899

Pe\~na-Cabrera G.V.Y \& Durand-Manterola H.J. (2004), Possible
biotic distribution in our galaxy, \emph{Advances in Space
Research}, {\bf{}33}, 114-117

Sackett, P.D. (1997), Does the Milky Way have a maximal disk?,
\emph{ApJ}, {\bf{}483}, 103-110

Spergel, D.N. \emph{et. al.} (2006), Wilkinson Microwave Anisotropy
Probe (WMAP) Three Year Results: Implications for Cosmology,
\emph{submitted to ApJ}, astro-ph/0603449

Tipler, F.J. (1980), Extraterrestrial intelligent beings do not
exist, \emph{Quarterly Journal of the Royal Astronomical Society},
{\bf{}21}, 267–281

Zheng, Z.; Flynn,C.; Gould,A.; Bahcall,J. N.; SALIM, S. (2001), M
Dwarfs from Hubble Space Telescope Star Counts , \emph{ApJ},
{\bf{}555}, 393-404

\end{list}
\end{document}